\documentclass{article}

\usepackage{natbib}
\usepackage{textcomp}
\setcitestyle{square}
\usepackage[hyphens]{url}
\usepackage{amsmath}
\usepackage{graphicx}
\graphicspath{{./figures/}}

\renewcommand{\thefigure}{\emph{\arabic{figure}}}
\usepackage[margin=0.5in, footskip=0.25in]{geometry}

\begin{document}

\noindent
\begin{minipage}{\textwidth}
\begin{flushleft}
\Large
\textbf{Satellite monitoring of annual US landfill methane emissions and trends}\\[1ex]
\large 
Nicholas Balasus\textsuperscript{a}, Daniel J.~Jacob\textsuperscript{a}, Gabriel Maxemin\textsuperscript{a}, Carrie Jenks\textsuperscript{b}, Hannah Nesser\textsuperscript{c},\\ Joannes D.~Maasakkers\textsuperscript{d}, Daniel H.~Cusworth\textsuperscript{e}, Tia R.~Scarpelli\textsuperscript{e}, Daniel J. Varon\textsuperscript{a}, Xiaolin Wang\textsuperscript{a}\\[1ex]
\footnotesize
\textsuperscript{a}\emph{School of Engineering and Applied Sciences, Harvard University, Cambridge, MA 02138.}\\
\textsuperscript{b}\emph{Environmental and Energy Law Program, Harvard University, Cambridge, MA 02138.}\\
\textsuperscript{c}\emph{Jet Propulsion Laboratory, California Institute of Technology, Pasadena, CA 91011.}\\
\textsuperscript{d}\emph{SRON Netherlands Institute for Space Research, Leiden, Netherlands.}\\
\textsuperscript{e}\emph{Carbon Mapper, Pasadena, CA 91011.}\\
\end{flushleft}
\end{minipage}

\section*{Significance Statement}
Landfills are estimated to be the third largest anthropogenic methane emission source in the United States. Individual landfills report their emissions to the US Environmental Protection Agency (EPA), which uses them in national reports tracking US progress in mitigating greenhouse gas emissions. We use satellite observations of methane to quantify annual 2019--2023 emissions from four US landfills. We find emissions to be 6$\times$ higher than reported to the EPA, and increasing rather than decreasing. Landfills have the option to calculate their emissions using either a generation-first or recovery-first model. They prefer the recovery-first model which gives lower emissions. The generation-first model is more consistent with atmospheric observations and implies that US landfill emissions are increasing instead of decreasing.

\section*{Abstract}
We use satellite observations of atmospheric methane from the TROPOMI instrument to estimate total annual methane emissions for 2019--2023 from four large Southeast US landfills with gas collection and control systems. The emissions are on average 6$\times$ higher than the values reported by the landfills to the US Greenhouse Gas Reporting Program (GHGRP) which are used by the US Environmental Protection Agency (EPA) for its national Greenhouse Gas Inventory (GHGI). We find increasing emissions over the 2019--2023 period whereas the GHGRP reports a decrease. The GHGRP requires gas-collecting landfills to estimate their annual emissions either with a recovery-first model (estimating emissions as a function of methane recovered) or a generation-first model (estimating emissions from a first-order decay applied to waste-in-place). All four landfills choose to use the recovery-first model, which yields emissions that are one-quarter of those from the generation-first model and decreasing over 2019--2023, in contrast with the TROPOMI observations. Our TROPOMI estimates for two of the landfills agree with the generation-first model, with increasing emissions over 2019–2023 due to increasing waste-in-place or decreasing methane recovery, and are still higher than the generation-first model for the other two landfills. Further examination of the GHGRP emissions from all reporting landfills in the US shows that the 19\% decrease in landfill emissions reported by the GHGI over 2005--2022 reflects an increasing preference for the recovery-first model by the reporting landfills, rather than an actual emission decrease. The generation-first model would imply an increase in landfill emissions over 2013--2022, and this is more consistent with atmospheric observations.

\section*{Introduction}

Landfills generate methane by providing anaerobic conditions for organic waste decomposition. Methane is a powerful greenhouse gas, and reducing methane emissions is a priority for mitigating near-term climate change \citep{shindell2024}. The Greenhouse Gas Inventory (GHGI) of the US Environmental Protection Agency (EPA) estimates that landfills account for 17\% of anthropogenic methane emissions in the US \citep{epa2024}. This estimate is based on bottom-up methods relating landfill emissions to waste-in-place and gas capture. However, aircraft and satellite observations suggest that actual landfill emissions in the US are about 50\% higher than the GHGI \citep{cusworth2024, nesser2024}. Here, we use oversampling of 2019--2023 Tropospheric Monitoring Instrument (TROPOMI) satellite observations to infer annual emissions and year-to-year trends of selected landfills, and we examine the causes of differences with bottom-up estimates.

Large municipal solid waste (MSW) US landfills generating more than 25000 metric tons of CO$_2$-equivalents must report their annual methane emissions following standard bottom-up protocols to the EPA Greenhouse Gas Reporting Program (GHGRP) \citep{cfr40p98}. The GHGRP numbers are used by the GHGI to report total US landfill emissions to the United Nations Framework Convention on Climate Change under the Paris Agreement \citep{unfccc2015}. 1123 MSW landfills reported to the GHGRP in 2022, to which the GHGI added 11\% to account for non-reporting MSW landfills and 0.7 Tg a$^{-1}$ for industrial landfills, resulting in a total emission of 4.3 Tg a$^{-1}$ from US landfills reported to the UNFCCC. Again using GHGRP numbers, the EPA GHGI reports a 19\% decrease in methane emissions from landfills from 2005 to 2022 \citep{epa2024}.

TROPOMI is uniquely suited to monitor total annual emissions from individual landfills and their trends. Unlike instruments that observe methane enhancements from point sources \citep{cusworth2024,dogniaux2024}, TROPOMI measures total column methane, allowing for quantification of total emissions over the landfill area. In addition, TROPOMI provides daily global mapping in contrast to the more limited snapshots from point source instruments. TROPOMI emission estimates are thus more directly comparable to the annual emissions reported to the GHGRP. TROPOMI has continuous observations since April 2018, allowing for multiple years of emissions estimates for trend analysis. The spatial resolution (5.5 $\times$ 7 km$^2$ at nadir) can be effectively increased to 1 $\times$ 1 km$^2$ by oversampling multiple days of observations \citep{maasakkers2022}, allowing for the isolation of landfill plumes under favorable conditions (successful observations, spatial separation from other emissions sources).

\section*{Results}

We use wind-rotated oversampling of TROPOMI observations for each year from 2019 to 2023 to construct annually averaged methane plumes with 1 $\times$ 1 km$^2$ resolution from four large Southeast US landfills (\emph{Materials and Methods}): Sampson County Disposal Landfill (Roseboro, North Carolina), Charlotte Motor Speedway Landfill (Concord, North Carolina), Lee County Landfill (Bishopville, South Carolina), and Springhill Regional Landfill (Campbellton, Florida). We chose these four landfills for their isolation from other sources and a high density of successful TROPOMI retrievals in all seasons and for each year. Fig.~1 shows the plumes for individual years. Each includes 95--129 days of successful observations. We applied the cross-sectional flux method \citep{varon2018} to quantify total annual emissions from the individual landfills and associated uncertainties (\emph{Materials and Methods}).

We estimate mean 2019--2023 emissions of 30 $\pm$ 3 Gg a$^{-1}$ from Sampson County Disposal Landfill, which can be compared to observed instantaneous emissions of 41 $\pm$ 17 Gg a$^{-1}$ (N = 1) with the AVIRIS-NG aircraft instrument in 2022 \citep{cusworth2024}, 47 $\pm$ 9 Gg a$^{-1}$ (N = 3) from individually-detected TROPOMI plumes in 2021 and 2022 \citep{schuit2023,dogniaux2024}, and 7 $\pm$ 4 Gg a$^{-1}$ (N = 9) from GHGSat plumes in 2021 and 2022 \citep{dogniaux2024}. We estimate 29 $\pm$ 2 Gg a$^{-1}$ from Charlotte Motor Speedway Landfill, while the same previous studies of instantaneous emissions report 25 $\pm$ 7 Gg a$^{-1}$ (N = 3) from AVIRIS-NG, 103 $\pm$ 52 Gg a$^{-1}$ (N = 18) from TROPOMI, and 20 $\pm$ 5 Gg a$^{-1}$ (N = 5) from GHGSat. No previous top-down estimates have been reported for Lee County and Springhill Regional landfills, for which we estimate mean 2019--2023 emissions of 28 $\pm$ 4 Gg a$^{-1}$ and 29 $\pm$ 4 Gg a$^{-1}$ respectively. Previous studies could have a biased representation of annual emissions due to infrequent sampling, reliance on detectable plumes (no null observations), and/or omission of diffuse area emission contributions. Our results are consistent with \citet{nesser2024} who estimated total annual US emissions at up to 25-km resolution for 2019 by Bayesian inversion of TROPOMI observations and reported 25 (23–29) Gg a$^{-1}$ for Sampson County Disposal Landfill and 23 (18–30) Gg a$^{-1}$ for Charlotte Motor Speedway Landfill.

We find that the emissions from these four landfills are on average 6$\times$ higher than reported to the GHGRP following the EPA protocol. Because these four landfills all have gas collection and control systems, they are required by the EPA to calculate their emissions using both a recovery-first and a generation-first model \citep{cfr40p98}. They can then select the result that they believe best represents their emissions [40 CFR \S 98.346(i)(13)]. The generation-first and recovery-first models correspond to Equations HH-6 and HH-8 in 40 CFR Part 98 Subpart HH respectively. Both models start from Equation (1):
\begin{gather}
	E = (G - R)\times(1-OX) + ID \label{eqn1}
\end{gather}
where $E$ is the methane emitted for the reporting year, $G$ is the methane generated inside the landfill, and $R$ is the methane recovered with the gas collection and control system. Any methane that is generated but not recovered exits through the landfill cover where a fraction, $OX$, is oxidized. $ID$ accounts for emissions from methane recovered but not fully destroyed (such as by an inefficient flare). The difference between the generation-first and recovery-first models lies in how $G$ is calculated. In the generation-first model, $G$ is based on first-order-decay of the waste deposited in the landfill. In the recovery-first model, $G$ is based on taking the measured amount of methane recovered and assuming it was recovered with a certain efficiency. This efficiency is parameterized based on cover type and gas collection system coverage, and averaged 72\% across GHGRP-reporting landfills in 2022. $OX$ is parameterized based on the surface area of the landfill and the difference between $G$ and $R$, and ranged from 0 to 35\% across GHGRP-reporting landfills in 2022. 

Emission estimates from the generation-first and recovery-first models for individual years in 2019--2022 are shown in Fig.~2 alongside our TROPOMI-based emission estimates. GHGRP values are not yet available for 2023 as of this writing. The recovery-first model yields emissions one-quarter of those from the generation-first model and shows flat or declining trends over the 2019--2022 period whereas the generation-first model shows increasing trends. Over the 2019--2022 period, all four landfills consistently chose the recovery-first model and its lower emissions total. This choice would not be supported by our emission estimates. For Sampson County Disposal and Charlotte Motor Speedway, the generation-first model and its increasing trend align well with our estimates. According to the generation-first model, the emissions from these two landfills are increasing because they are maintaining or increasing their rate of waste deposition while recovering less methane (Fig.~S1--S2). The recovery-first model may erroneously report decreasing emissions as it assumes that less methane recovered means that emissions have decreased. However, less methane recovery could alternatively indicate a recovery system malfunction or inefficiency.

Our TROPOMI emission estimates for Lee County and Springhill Regional are higher than either of the models. At Lee County, the generation-first model shows a consistent increase. The spike in emissions in 2020 could be a result of abnormally high rainfall (Fig.~S3) that could impact the effectiveness of the gas collection system \citep{cusworth2024} and alter the waste decay rate \citep{wang2024}, though Sampson County Disposal and Charlotte Motor Speedway also experienced similar rainfall. The gas collection system at Lee County expanded from 209 to 214 wells in 2020, a process that involves drilling into the landfill, which can produce large emissions \citep{cusworth2020}, though the other landfills also expanded their gas collection systems. Landfill bottom-up models may struggle to capture these types of events \citep{cusworth2024,nesser2024}. Our estimated emissions from Springhill Regional decreased over 2019--2023, potentially driven by decreasing waste deposition (Fig.~S4), though methane recovery decreased as well. There was construction at this landfill (186 wells in 2019 expanding to 201 wells in 2022) that may again contribute to emissions not captured by the generation-first model. Shortcomings of the generation-first model applied to Lee County and Springhill Regional could also be related to landfill malfunctions, such as cracks in the landfill cover or leaks in the recovery system \citep{cusworth2024,nesser2024}.

\section*{Discussion}

TROPOMI observations of annual 2019--2023 methane emissions from four US landfills suggest that the current protocol for facilities to report emissions to the US GHGI national inventory through the GHGRP may cause severe underestimation of emissions. The recovery-first model used by all four landfills appears to underestimate emissions and shows erroneous decreasing trends. The generation-first model matches our TROPOMI-based estimate for two of the landfills and features increasing trends, but the other two landfills are still lower. Previous point-source observations from GHGSat suggested that the recovery-first model is an underestimate and the generation-first model is an overestimate \citep{dogniaux2024} but these observations could miss the diffuse area emissions and cover much fewer days per year compared to our study.

Of the 1123 MSW landfills reporting to the GHGRP in 2022, 283 did not have a gas collection system and had to use the generation-first model. 840 landfills had gas collection systems and had the flexibility to report their emissions either with the recovery-first or generation-first model. Of these, 570 landfills selected the recovery-first model and 219 landfills selected the generation-first model. 49 landfills recovered more methane than produced from the generation-first model and thus were required to use the recovery-first model. 2 landfills reported emissions inconsistent with either model.

70\% of MSW landfills thus had the option to pick their emissions calculation model as of 2022. To show the resulting uncertainty and similarly to \citet{stark2024}, we calculated the total national landfill emissions for two different scenarios where all landfills with flexibility are required to choose one model or the other (Fig.~3). Requiring the generation-first model increases total US landfill emissions by almost a factor of two in 2022 compared to requiring the recovery-first model. The EPA GHGI reports total US landfill emissions of 4.3 Tg a$^{-1}$, which is closer to the recovery-first scenario, but our work and previous top-down studies constraining the national total [Lu et al., 2022, using GOSAT satellite and NOAA surface data; Nesser et al., 2024, using TROPOMI data] are closer to the generation-first scenario. The GHGI reports a 19\% decrease in US landfill emissions from 2005 to 2022, with 2005 being the base year for the commitments of the US to the Paris Agreement \citep{usa2021} We find that this decreasing trend is driven by the increasing preference of landfills to use the recovery-first model in their reporting. Stated more generally, as shown on the right in Fig.~3, when presented with the option of selecting either the recovery-first or generation-first emission estimate in their reporting, landfills are increasingly choosing the lower emission value. In contrast to the GHGI, the generation-first model suggests increasing emissions from 2013 that total 6.7 Tg a$^{-1}$ in 2022, matching the corresponding estimate for 2005 and implying no decrease in emissions. As shown in Fig.~S5, the increasing trend in emissions over the past decade is supported by waste deposition being at an all-time high (up 30\% from 2010 to 2022), which methane recovery has not kept up with (up only 6\% from 2010 to 2022).

In summary, TROPOMI satellite observations of 2019--2023 annual emissions for four US gas-collecting landfills show emissions 6$\times$ higher than reported to the Greenhouse Gas Reporting Program (GHGRP) and used in the US national Greenhouse Gas Inventory (GHGI) to report total landfill emissions. These four landfills have the option of using either a recovery-first or a generation-first model in reporting their emissions to the GHGRP. They all choose the recovery-first model, which is considerably lower than the generation-first model and shows decreasing trends inconsistent with the TROPOMI observations. Examination of the ensemble of US landfills shows that the low GHGI emissions and 19\% decrease over 2005--2022 are a result of an increasing percentage of gas-collecting landfills choosing to report the low recovery-first emissions as opposed to the generally higher generation-first emissions. Recent amendments to the GHGRP could resolve shortcomings of (1) the generation-first model by increasing decay rates, (2) the recovery-first model by decreasing assumed collection efficiencies, and (3) both models by encouraging accounting for issues with destruction devices and recovery systems \citep{89fr31802}. However, the large discrepancy between the two models will likely persist, and the models may still struggle with emissions from the working face or from construction events. Bias in the EPA GHGI and its trend are of relevance for the Nationally Determined Contribution of the US to the Paris Agreement to reduce GHG emissions by 50--52\% by 2030 relative to 2005 \citep{usa2021}. Comprehensive monitoring is necessary to verify the effectiveness of emissions reduction policies, including EPA’s planned update to its Clean Air Act regulations of emissions from new and existing landfills \citep{whitehouse2024}.

\section*{Materials and Methods}
\subsection*{Satellite Observations}

The Tropospheric Monitoring Instrument (TROPOMI), onboard the ESA Sentinel-5 Precursor low-Earth orbiting satellite, observes atmospheric methane concentrations with continuous global daily coverage at 13:30 local time and 5.5 $\times$ 7 km$^2$ nadir spatial resolution \citep{lorente2021}. The observations are by solar backscatter in the 2.3 $\mu$m methane absorption band. Successful retrievals are limited by cloud cover and surface reflectivity anomalies \citep{butz2012}. The methane column retrieved from the satellite radiances is converted to a column average dry air mixing ratio, denoted XCH$_4$, in units of parts per billion (ppb). We use the blended TROPOMI+GOSAT product of \citet{balasus2023}, which corrects TROPOMI artifacts by using the more precise but much sparser GOSAT satellite data \citep{parker2020}. We further remove residual striping in the data following \citet{borsdorff2024}.

\subsection*{Emissions Calculation}

We choose four US landfills that are isolated from other methane sources as indicated by the gridded GHGI \citep{maasakkers2023} and are successfully observed by TROPOMI in all seasons and for each year. We aim to quantify the annual emission rate from each landfill so we average the TROPOMI data across time, sacrificing the time resolution to effectively increase the spatial resolution. This technique is known as oversampling and takes advantage of shifts in satellite orbits from day to day resulting in partial overlap between pixels \citep{zhu2017}. We oversample one year of data at a time, using a tessellation approach that averages the XCH$_4$ footprints to a 0.6\textdegree{} $\times$ 0.6\textdegree{} grid at 0.01\textdegree{} resolution centered on the landfill. Before oversampling, we exclude XCH$_4$ observations further than 3 standard deviations from the mean XCH$_4$ in the 0.6\textdegree{} $\times$ 0.6\textdegree{} scene. We also associate each TROPOMI observation with a wind vector and rotate the observations around the landfill to align this vector in the westerly direction \citep{maasakkers2022}. We collect hourly zonal and meridional winds at 3 km resolution from the High-Resolution Rapid Refresh model (HRRR) \citep{dowell2022} and perform a pressure-weighted average over the lowest 100 hPa of the atmosphere \citep{potts2023}. We obtain hourly zonal and meridional winds at the landfill locations by averaging together the resulting winds from HRRR grid cells within 5 km. The wind vector associated with each satellite observation is the average of these landfill winds over the three hours before the satellite observation time.

Following wind rotation and oversampling, we now have time-averaged plumes of XCH$_4$ for each year. An emission rate can be inferred from the methane mass enhancement and wind speed \citep{jacob2022}. We use the cross-sectional flux method \citep{frankenberg2016,varon2018} to perform our emissions estimates, as has been done with wind-rotated TROPOMI observations before \citep{schneising2020,schneising2024}.

For our best estimate of the annual emissions, working from the oversampled scenes in Fig.~1, we use ten north-south transects centered at the landfill latitude and 0.2\textdegree{} in length, starting just downwind of the landfill location and moving downwind with 0.01\textdegree{} spacing. The mean oversampled XCH$_4$ outside this region is defined as the background XCH$_4$. Subtracting the background XCH$_4$ from all values gives a concentration enhancement. Multiplying by dry air columns (available in the TROPOMI product and oversampled in the same manner as XCH$_4$) leads to a mass enhancement $\Delta\Omega$ (kg m$^{-2}$). Integrating the mass enhancement in the crosswind direction $y$ between bounds $a$ and $b$ (defined by the 0.2\textdegree{} length) and multiplying by annual average winds $U$ (m s$^{-1}$) gives the emission rate Q (kg s$^{-1}$) at each 0.01\textdegree{} distance $x$ downwind of the source:
\begin{gather}
Q = U\int_{a}^{b}\Delta\Omega(x,y)\,dy
\end{gather}

The annual average winds are defined as the harmonic mean of the wind speeds associated with each satellite observation used in the oversampling for that year. Since the methane enhancement $\Delta\Omega$ is proportional to the ratio of the emission rate to wind speed ($\Delta\Omega \sim Q/U$), it is most appropriate to use harmonic means for wind speeds when interpreting arithmetic mean methane enhancements. We discard scenes with $U$ $<$ 1 m s$^{-1}$ because of large errors in applying the cross-sectional
flux method to such weak winds \citep{varon2018}.

Equation (2) results in ten different emission estimates, one for each transect. We average them together to get the best estimate. We estimate uncertainty with a bootstrapping method, repeating our emission estimation process ten thousand times. Each time, we sample with replacement the arrays used for calculating background XCH$_4$, the annual wind speed, and the mean emission estimate across the ten transects. The result is a distribution of ten thousand estimates of $Q$, from which we take the 2.5$^{\text{th}}$ and 97.5$^{\text{th}}$ percentile to define the 95\% confidence intervals as our reported uncertainty.

\subsection*{EPA GHGRP Data}

We scrape data reported by the landfills to the EPA GHGRP from the FLIGHT tool (\url{http://ghgdata.epa.gov/ghgp/main.do}, accessed: 29 July 2024). These data are updated each year and landfills can revise their reports from previous years, so we archive our scraped results as a part of this study. For Fig.~3, the total sector emissions are calculated consistent with EPA GHGI methodology by summing the methane emissions reported by each MSW landfill to the GHGRP and scaling up by 9\% (2010--2016) or 11\% (2017--2022) to account for non-reporting MSW landfills. We also add emissions reported by industrial landfills to be consistent with \citet{lu2022} and \citet{nesser2024}, which we take from the GHGI \citep{epa2024}. In the alternative scenarios in Fig.~3, we only alter the reports of gas-collecting MSW landfills that are given the flexibility to pick either the recovery-first or generation-first model. 

\section*{Data, Materials, and Software Availability}
The blended TROPOMI+GOSAT data (\url{https://registry.opendata.aws/blended-tropomi-gosat-methane/}) and HRRR data (\url{https://registry.opendata.aws/noaa-hrrr-pds/}) are available through the AWS Open Data Sponsorship Program. The code for this study is available through GitHub (\url{https://github.com/nicholasbalasus/united_states_landfill_methane}) and will be stored on Zenodo after peer review. AVIRIS-NG data are available via Carbon Mapper's Open Data Portal (\url{https://data.carbonmapper.org}).

\section*{Acknowledgements}
This work was funded by the Harvard Methane Initiative and the NASA Carbon Monitoring System. Nicholas Balasus was supported by the National Defense Science and Engineering Graduate fellowship program from the Department of Defense. This research was also supported in part by an appointment to the NASA Postdoctoral Program at the Jet Propulsion Laboratory, California Institute of Technology, administered by Oak Ridge Associated Universities under contract with NASA. We thank the team that realized the TROPOMI instrument and its data products, consisting of the partnership between Airbus Defence and Space Netherlands, KNMI, SRON, and TNO and commissioned by NSO and ESA. The Sentinel-5 Precursor is part of the EU Copernicus program, and Copernicus (modified) Sentinel-5P data (2019--2023) have been used.

\clearpage
\bibliography{references.bib}

\clearpage
\begin{figure}[ht]
	\centering
	\includegraphics[width=\textwidth]{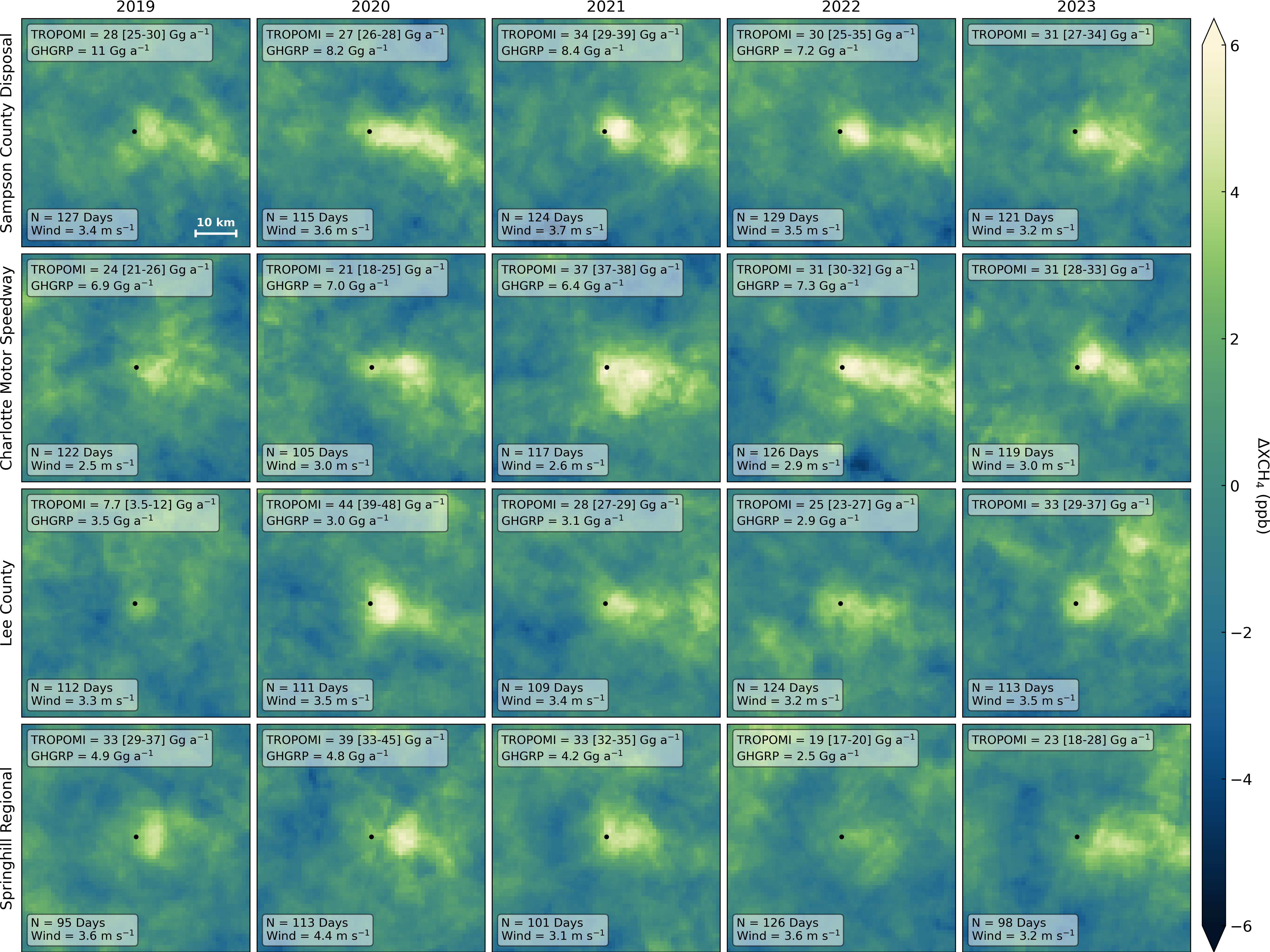}
	\caption{\emph{Annual time-averaged methane plumes from oversampled TROPOMI satellite observations for the Sampson County Disposal Landfill in North Carolina (GHGRP ID 1004118), the Charlotte Motor Speedway Landfill in North Carolina (1003099), the Lee County Landfill in South Carolina (1002557), and the Springhill Regional Landfill in Florida (1007898). Annual emissions derived from the TROPOMI observations with 95\% confidence intervals are compared to the values reported by the landfills to the US Greenhouse Gas Reporting Program (GHGRP). Also shown are the number of days when TROPOMI made successful observations in the scene along with the mean wind speeds used in the emissions calculations. TROPOMI observations of dry methane column mixing ratio (XCH\textsubscript{4}) are wind-rotated around the landfill (black dot) to simulate westerly winds and then oversampled on a 0.01\textdegree{} $\times$ 0.01\textdegree{} grid. Methane above background ($\Delta$XCH\textsubscript{4}) subtracts the background defined as the mean XCH\textsubscript{4} outside the 0.2\textdegree{} $\times$ 0.1\textdegree{} region downwind of the landfill (Materials and Methods). The scale is the same for all panels.}}
	\label{fig1}
\end{figure}

\begin{figure}[ht]
	\centering
	\includegraphics[width=\textwidth]{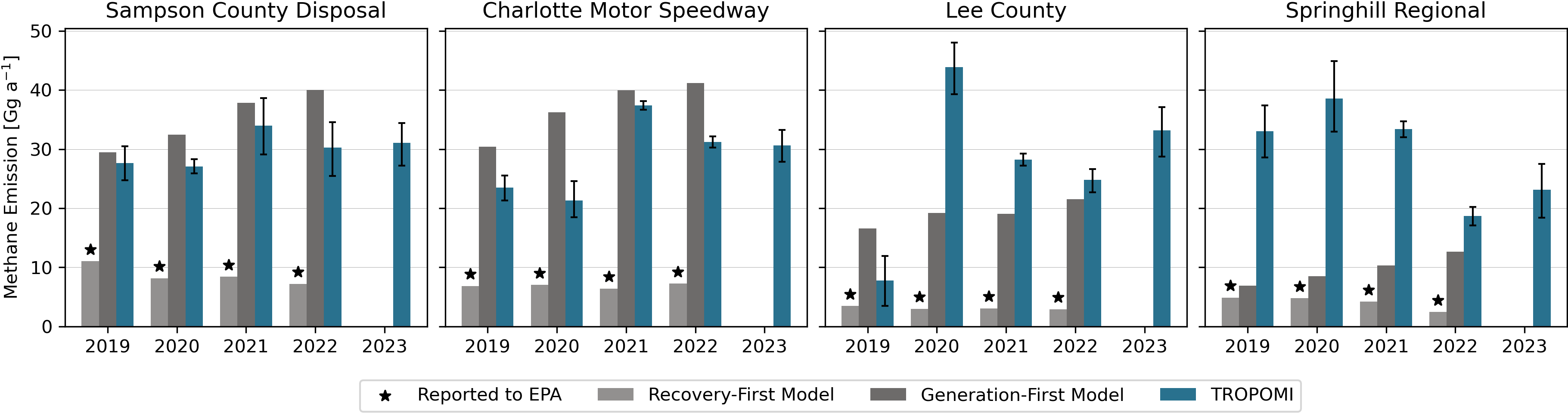}
	\caption{\emph{Estimates of annual methane emissions from four US landfills for the years 2019--2023. The first two bars for each year are the emissions calculated by the landfills using the recovery-first and generation-first models. The emissions they chose to report to the EPA for inclusion in the national inventory are indicated with a star. Reported emissions for 2023 are not yet available. The third bar for each year is the TROPOMI estimate with 95\% confidence intervals.}}
\end{figure}

\begin{figure}[ht]
	\centering
	\includegraphics[width=\textwidth]{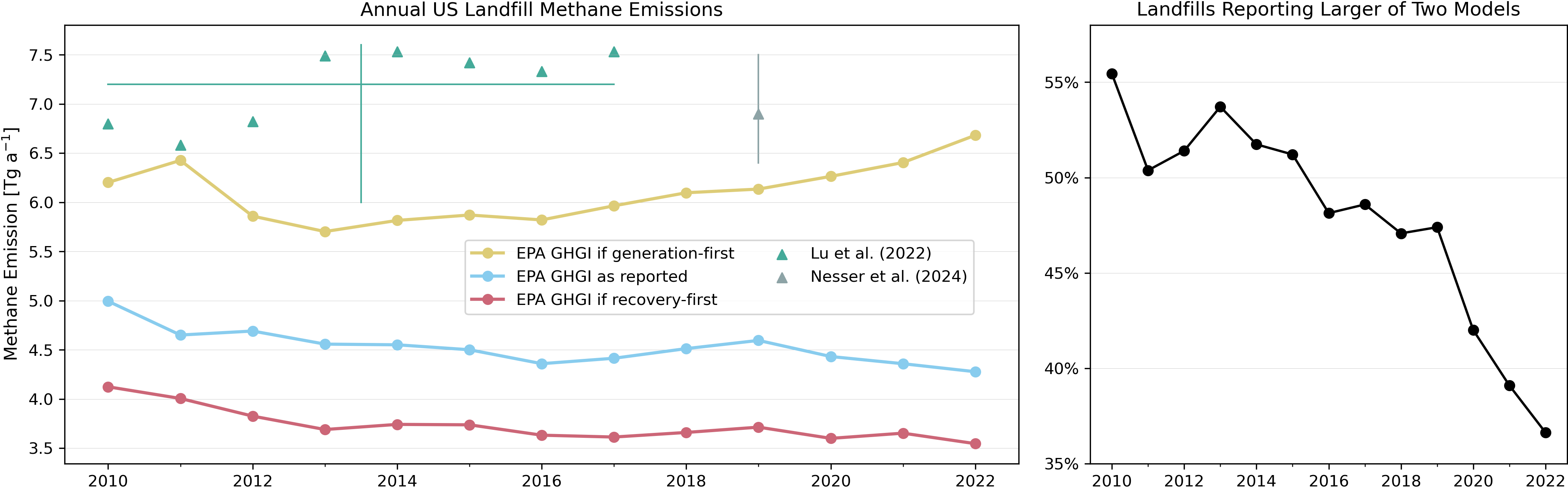}
	\caption{\emph{Magnitude and trend of US landfill emissions. The left panel shows annual US landfill methane emissions based on reports to the GHGRP by both municipal solid waste (MSW) and industrial landfills (Materials and Methods). The US EPA Greenhouse Gas Inventory (GHGI) sums national emissions from industrial landfills, non-gas-collecting MSW landfills, and gas-collecting MSW landfills that have selected either the generation-first or recovery-first model. The generation-first and recovery-first lines show what the GHGI emissions would be if all gas-collecting MSW landfills were required to choose that model for reporting. Top-down estimates of national landfill emissions are shown from \citet{lu2022} using GOSAT satellite and NOAA surface data and from \citet{nesser2024} using TROPOMI data. For \citet{lu2022}, the cross gives the 2010–2017 mean and its uncertainty. The right panel shows the evolution in the percentage of gas-collecting MSW landfills that opted to report the higher of the two models (generation-first or recovery-first) to the GHGRP.}}
\end{figure}

\clearpage
\section*{Supplementary Information}
\setcounter{figure}{0}
\renewcommand{\thefigure}{\emph{S\arabic{figure}}}

We compile variables related to the operating and meteorological conditions at the Sampson County Disposal Landfill (Fig.~S1), Charlotte Motor Speedway Landfill (Fig.~S2), Lee County Landfill (Fig.~S3), and Springhill Regional Landfill (Fig.~S4). The top row is values reported to the GHGRP by each landfill, including the amount of methane they recovered, their collection efficiency as parameterized based on the cover type, and the amount of waste deposited into the landfill. The bottom row is meteorological conditions from HRRR at the grid cell located over the landfill averaged over the entire year, including temperature, surface pressure, and precipitation. Temperature and surface pressure are taken from the 0$^{\text{th}}$ hour forecasts, while precipitation is taken as the precipitation accumulated in the previous hour from the 1$^{\text{st}}$ hour forecasts.

\begin{figure}[ht]
	\centering
	\includegraphics[width=0.8\textwidth]{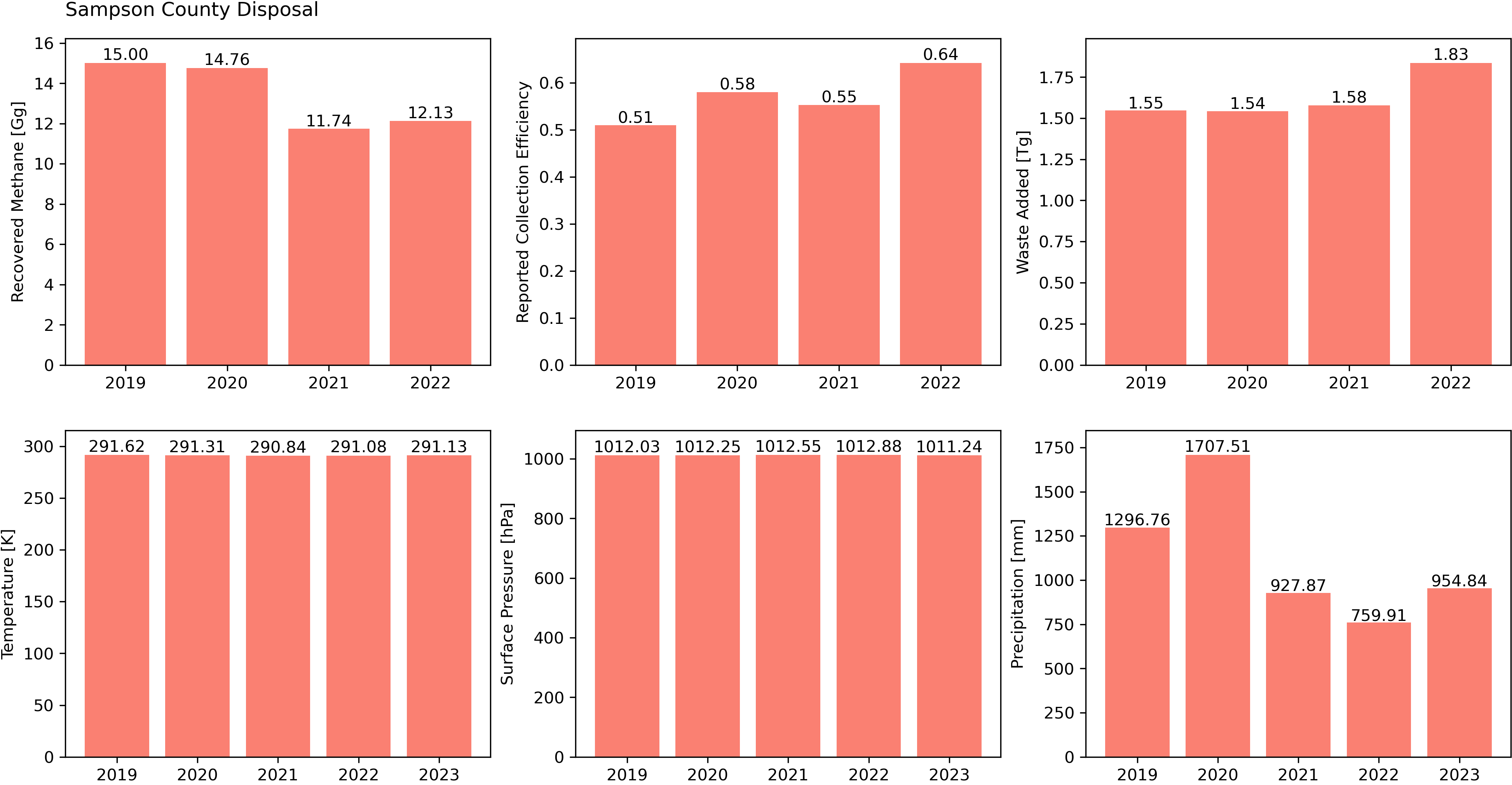}
	\caption{\emph{Potential driving variables of methane emissions from the Sampson County Disposal Landfill (34.983\textdegree{}N, 78.464\textdegree{}W) based on reports to the EPA GHGRP (top, 2019--2022) and HRRR meteorological data (bottom, 2019--2023).}}
\end{figure}

\begin{figure}[ht]
	\centering
	\includegraphics[width=0.8\textwidth]{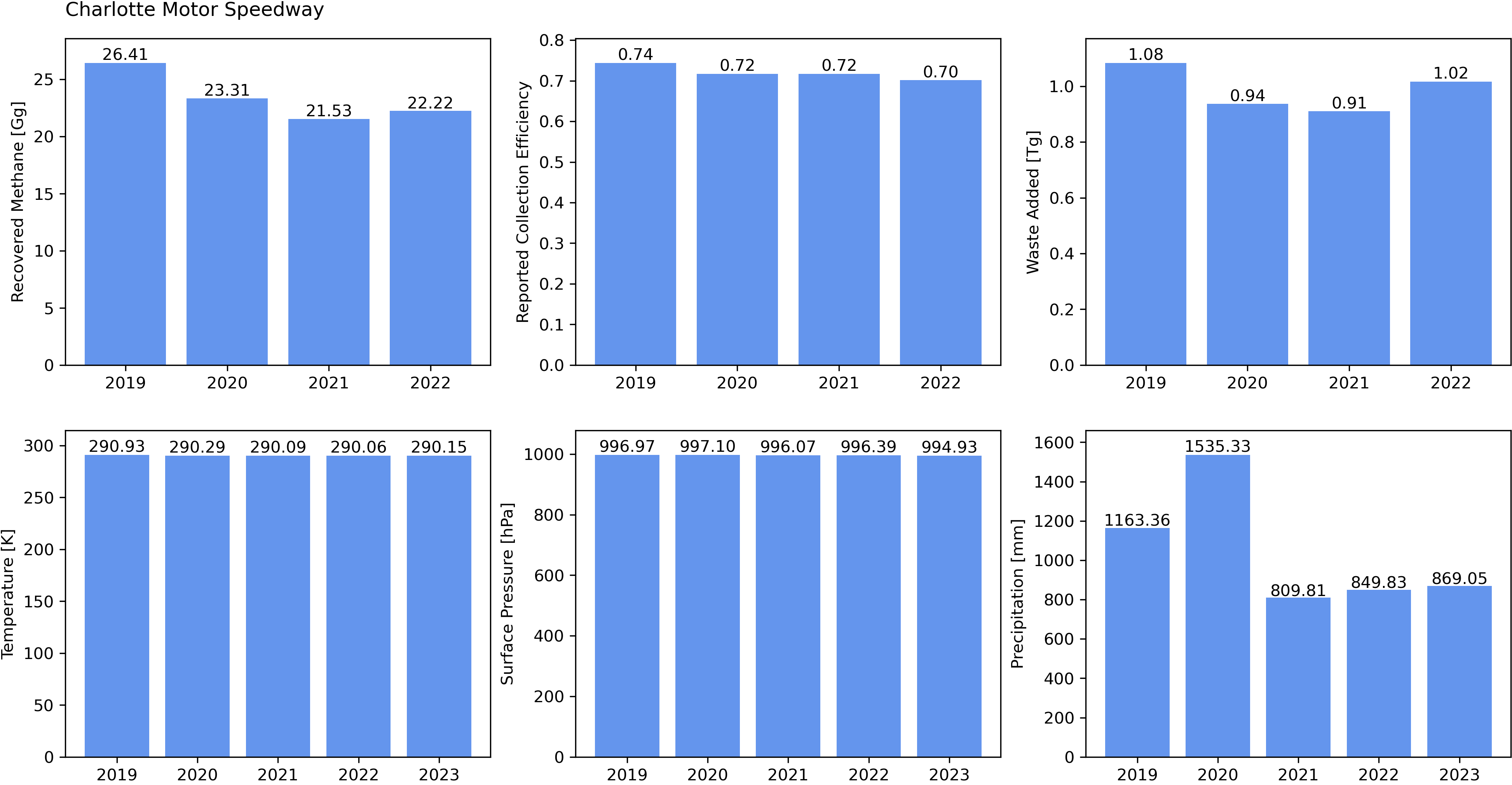}
	\caption{\emph{Potential driving variables of annual methane emissions from the Charlotte Motor Speedway Landfill (35.341\textdegree{}N, 80.658\textdegree{}W) based on reports to the EPA GHGRP (top, 2019--2022) and HRRR meteorological data (bottom, 2019--2023).}}
\end{figure}

\begin{figure}[ht]
	\centering
	\includegraphics[width=0.8\textwidth]{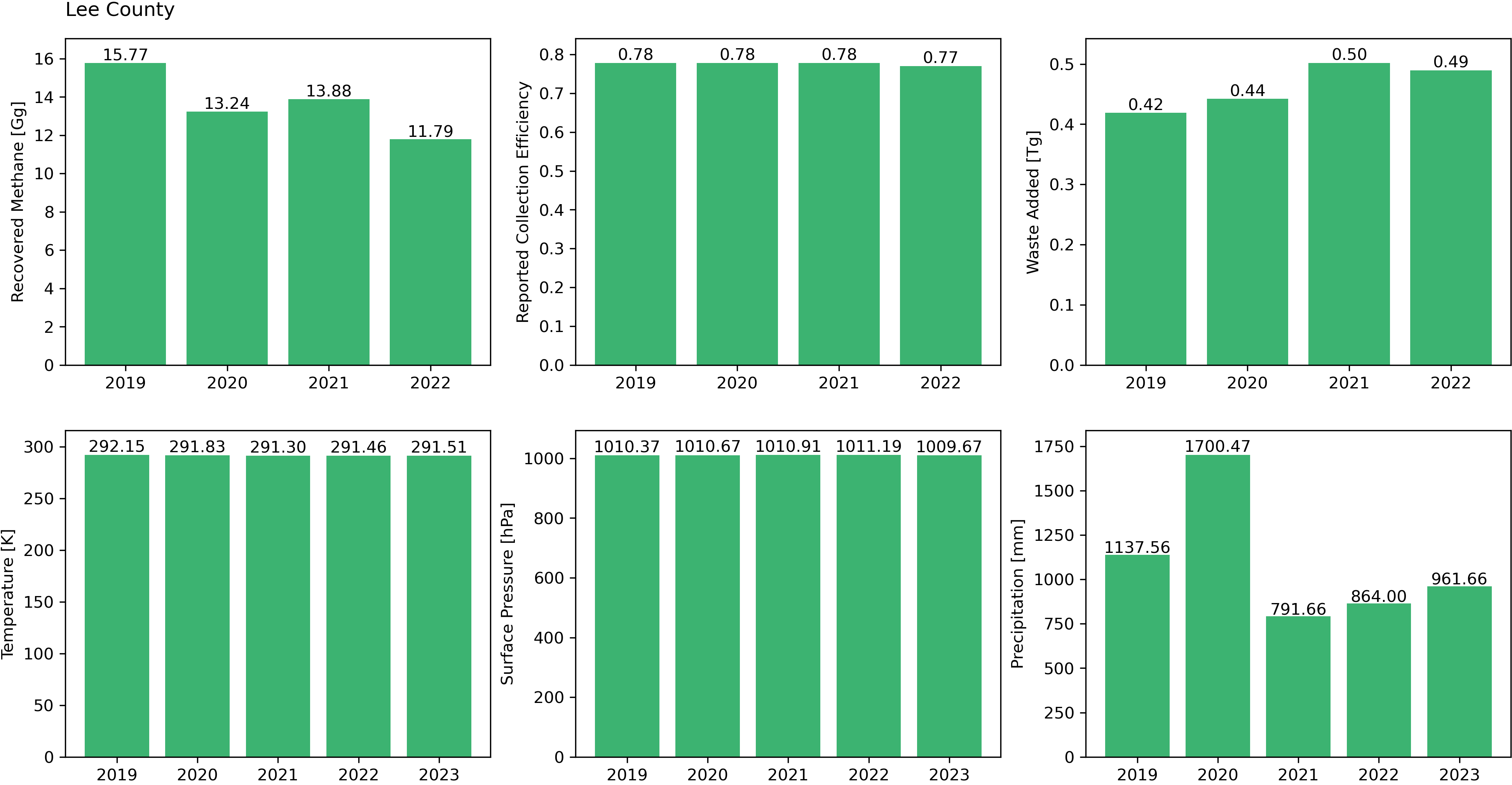}
	\caption{\emph{Potential driving variables of annual methane emissions from the Lee County Landfill (34.178\textdegree{}N, 80.272\textdegree{}W) based on reports to the EPA GHGRP (top, 2019--2022) and HRRR meteorological data (bottom, 2019--2023).}}
\end{figure}

\begin{figure}[ht]
	\centering
	\includegraphics[width=0.8\textwidth]{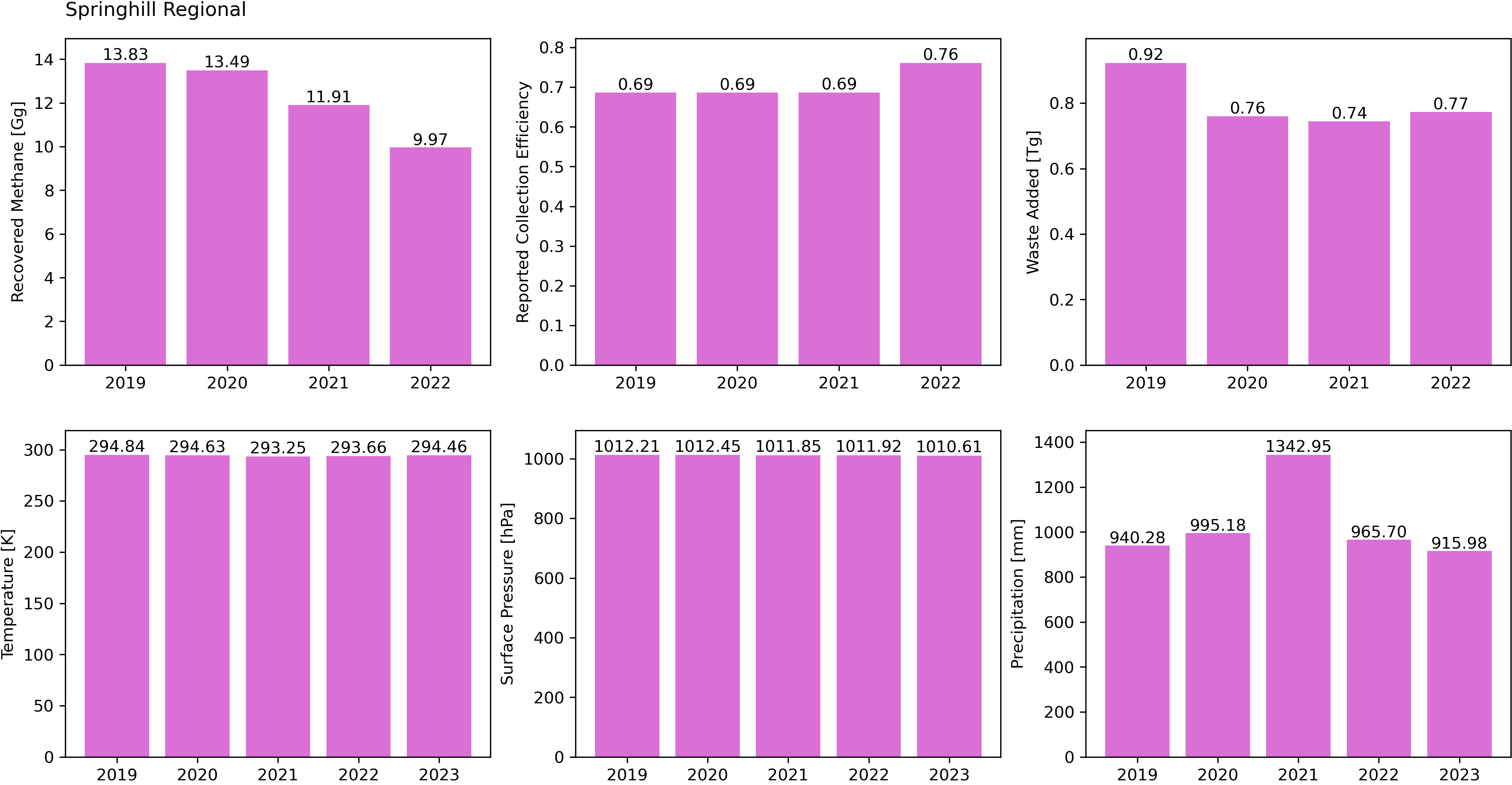}
	\caption{\emph{Potential driving variables of annual methane emissions from the Springhill Regional Landfill (30.923\textdegree{}, 85.439\textdegree{}W) based on reports to the EPA GHGRP (top, 2019--2022) and HRRR meteorological data (bottom, 2019--2023).}}
\end{figure}

\begin{figure}[ht]
	\centering
	\includegraphics[width=0.8\textwidth]{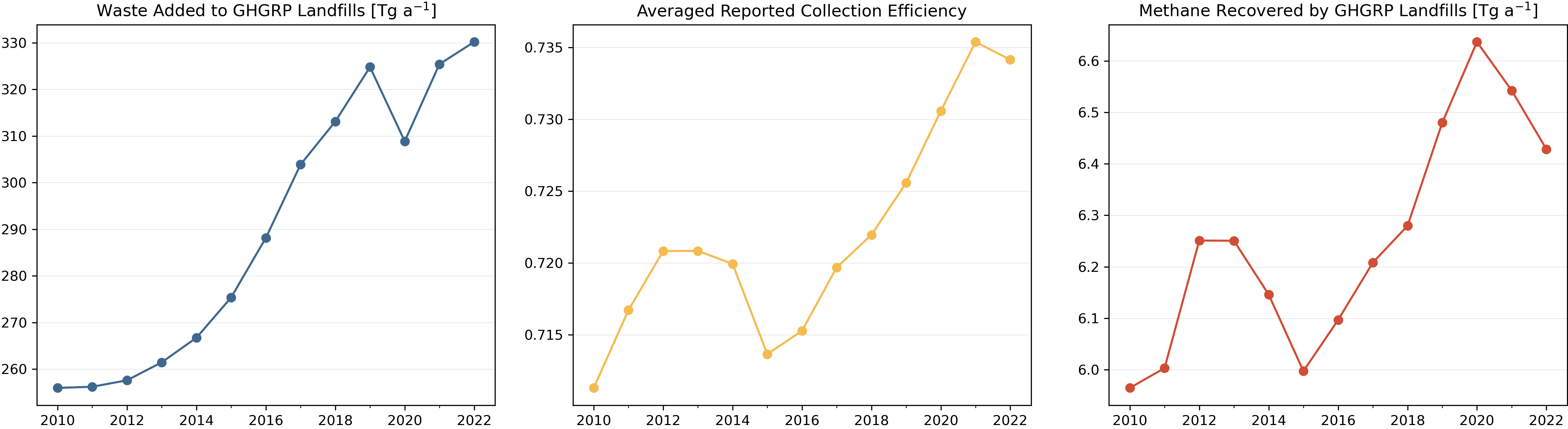}
	\caption{\emph{On the left, the amount of waste added to the landfills reporting to the GHGRP each year. In the middle, the average reported collection efficiency for landfills with gas collection and control systems reporting to the GHGRP. On the right, the amount of methane recovered by GHGRP landfills each year.}}
\end{figure}

\end{document}